 \documentstyle[11pt,epsf]{article}
 \newfont{\Bbb}{bbmss12}
 \newcommand{\N}{\mbox{\Bbb N}} 

\begin{document}

 \title{Clumps into Voids }
 \author{Nazeem Mustapha\thanks{email:~\tt nazeem@dyson.mth.uct.ac.za} and
Charles Hellaby\thanks{email:~\tt cwh@maths.uct.ac.za}}
 \maketitle
 \begin{center}
 Department of Mathematics and Applied Mathematics\\
 University of Cape Town\\
 Rondebosch \\
 7701
 \\[5mm]
 {\em Submitted to GRG 20/4/2000} \\[2mm]
 {\em astro-ph/0006083} \\[2mm]
 \end{center}

 \begin{abstract}
 We consider a spherically symmetric distribution of dust and show that it
is possible, under general physically reasonable conditions, for an
overdensity to evolve to an underdensity (and vice versa).  We find the
conditions under which this occurs and illustrate it on a class of regular
Lema\^{\i}tre-Tolman-Bondi ({\sc ltb}) solutions.  The existence of this
phenomenon, if verified, would have the result that the topology of
density contours, assumed fixed in standard structure formation theories,
would have to change and that luminous matter would not trace the dark
matter distribution so well.
 \end{abstract}
 \section{Introduction}
 In the inflationary universe paradigm, it is believed that the observed
universe is very nearly flat. The density of baryons --- which can be
obtained from primordial nucleosynthesis theory --- is however very small
and this requires that most matter is non-baryonic.  Traditional theories
of structure formation  assert that baryonic matter fell into the high
density peaks of dark matter and became luminous forming stars and
galaxies.
 The  stationary view, in which matter concentrations remain essentially
fixed, may be thought of as
being  governed by a mapping which preserves extremal points of
the density field. It may
well be a good approximation if the initial density field is simply
amplified by gravitational processing, but when the matter
content of the pre- and post-decoupling epochs is viewed from a
hydrodynamical point of view as a fluid in
 high-temperature plasma or
 quasi-plasma state, one would expect shock waves and other spatial
gradients to exist (even if their amplitudes were small).  Indeed, large
scale inhomogeneities and flows have been shown to be a pervasive
influence on the behaviour of the universe on scales of up to 100 Mpc.
(Cf. for example \cite{bi:P93,bi:C99,bi:HSLSD99}).
The now undisputed existence of
 large-scale cosmic flows
(on the scale of $15 000~km s^{-1}$) as has been
reported by various authors \cite{bi:RC88}, lends more
credence to the idea that perhaps the
stationary approximation, used ubiquitously in structure formation,
is not as good as assumed.  The bulk flow reading of
$700 \pm 170 km s^{-1}$ found for all Abell clusters with redshifts less
than $15 000 km s^{-1}$ strongly excludes any of the popular models with
Gaussian initial conditions.

In this context, the Lema\^{\i}tre-Tolman-Bondi ({\sc ltb})
 \cite{bi:L31,bi:T34,bi:B47} universe is interesting as one may
analytically study the evolution of spherically symmetric inhomogeneities.
The discovery of large scale voids and walls in the eighties
sparked interest in the {\sc ltb} model as a means of investigating these, and
other similar, structures (for example
 \cite{bi:OVV78,bi:MSS83,bi:S84}).

The nonlinear
effects of large scale clumps (for example \cite{bi:RT81}) and voids
(\cite{bi:AFMS93,bi:S84}) on the production of anisotropies in the {\sc cmb}
has been studied using {\sc ltb} models numerically. The results have been
that a large part of the temperature anisotropies in the background
radiation (the dipole component) may be completely due to large scale
structures, but leave open the origin of other sources (for example quadrupole)
as truly cosmological. Also worthy of mention is the work done by Lake and
Pim \cite{bi:LP85, bi:PL86}.

These studies
concentrated on the description and feasibility of spherical inhomogeneities,
and were not too concerned with determining under what conditions
structures could change radically with evolution. Here we intend to initiate
analytical studies on this topic.

At the centre of symmetry of an {\sc ltb} universe, we must generically have a
position of extreme density. Thus at
the centre it is
not feasible to study the question of density waves, {\em per se},
since a wave is
defined by the fact that a maximum (or minimum) moves at some velocity away
from the worldline.
But at the centre we can ask the question: `{\em under what
conditions will a density maximum evolve into a density minimum or vice
versa?}'.
This is a first step towards a study of
cosmic flows in this model, since if this question can  be answered in the
affirmative, then it would naturally follow that in some region around the
centre over the time elapsed a maximum (or minimum) has to be traveling
away from the centre.

 If physical, this would raise questions about the validity of the
standard model of structure formation. It is particularly important to
{\sc cobe} analyses where the data (for example, hot spots and cold spots
corresponding to under- and
 over-densities respectively)
on the last scattering surface is `transferred' to the current epoch by use
of a function which does not assume that the peaks in the matter
distribution may change to troughs.
 \section{Preliminaries and Programme}
 \label{sec:descr}
 We are interested in whether the profile of a density inhomogeneity can
change significantly with evolution. Specifically, we want to know whether
a central maximum in density can evolve into a central minimum, or vice
versa.  For our investigation we use the simplest inhomogeneous
cosmological solution to the  Einstein field equations, the {\sc ltb}
model.

 This universe model is spherically symmetric, but in general radially
inhomogeneous.
 Space-time is described by a
 four-dimensional continuum filled by an irrotational  perfect fluid with
a dust equation of state.  We may choose the natural coordinate system
labelled by $\{x^{a}\}_{a = 0}^{3} = \{t,r,\theta,\phi\}$ suggested by the
spherical symmetry. The coordinates are assumed to be comoving with the
particles.  This allows a  definition of  a fluid velocity
$\displaystyle{u^a =
 \frac{dx^a}{dt}}$ such that $u^a = \delta^{a}_{0}$ and $u_au^a = -1$,
 which would mean that time coordinate $t$ is also cosmic time. For an
ideal fluid with mass density $\rho$ and vanishing pressure (dust), the
 energy-momentum tensor has the form $\displaystyle{T^{a b} = \rho u^{a}
u^{b}}$. The conservation of
 energy-momentum ${T^{ab}}_{;b} = 0$ confines  the dust to geodesics and
also implies that the mass of any portion of the fluid is conserved
through the twice contracted Bianchi Identities .

The metric in synchronous comoving coordinates can be written as
 \begin{equation}
ds^2 = -dt^2 + \frac{(R')^2}{1 + 2E}dr^2 + R^2(d\theta^2 + \sin^2\theta
d\phi^2)
 \label{eq:metric}
 \end{equation}
where $R = R(t,r,)$  acts as a transverse scale factor for individual comoving
particles, and $E = E(r)$ is an arbitrary function of the
integration which has a dynamic as well as a metric geometric role.
$R$ is also the areal radius, that is  $4 \pi R^2$ describes the surface
area of the
sphere at comoving radius $r$ at any time $t$ and thus $R(t,r) \geq 0$.

The expression for the invariant energy density, $\rho = \rho(t,r)$,
is obtained from the {\small{$tt$}} field equation:
 \begin{equation}
 8\pi\rho = \frac{2\,M'}{R^{2} \hspace*{.1cm} R'}
 \label{eq:density}
 \end{equation}
where $M=M(r)$ is another arbitrary function.

The {\small{$rr$}}, {\small{$\theta\theta$}} and {\small{$\phi\phi$}}
components of the {\sc efe} reduce to
the single equation of motion
 \begin{equation}
{1 \over 2} \left( \frac{\dot{R}}{R} \right)^2 = \frac{M}{R^3} + \frac{E}{R^2} \,.
 \label{eq:motion}
 \end{equation}

We define a `scale radius', $p(r)$, and `scale time', $q(r)$, for
 non-parabolic models as follows:
 \begin{equation}
p(r) = \frac{M}{\pm  E}
 \label{eq:p}
 \end{equation}
and
 \begin{equation}
q(r) = \frac{M}{\sqrt{\pm \, (2E)^3}}
 \label{eq:q}
 \end{equation}
 which may be viewed as alternative variables to $M$ and $E$. We (for
convenience) may sometimes mix these four variables in the equations
below.  The `\,+\,' is applicable in hyperbolic models and the `\,$-$\,'
is used for elliptic  models. In a
 re-collapsing model, the areal radius at maximum expansion is given by
$p(r)$  and the time from creation to destruction is $\pi \, q(r)$.

 The
 Friedmann-like equation can be solved (see for example \cite{bi:B47}) in
terms of parameter $\eta = \eta(t,r)$.  For a
 non-empty universe, $M \neq 0$, there are three solutions to
(\ref{eq:motion}):
 \begin{equation}
   R = \frac{p}{2} \phi_0  , \hspace*{1cm}  \xi = \frac{2\,(t - t_B)}{q}
 \label{eq:R2}
 \end{equation}
 where
 \begin{equation}
   \phi_0 = \left\{
      \begin{array}{l}
               \cosh(\eta) - 1, \\
               (1/2) \eta^2,    \\
               1 - \cos(\eta),
      \end{array}
             \right.
   ~~~ \xi = \left\{
         \begin{array}{ll}
               \sinh(\eta) - \eta,  & \hspace*{1cm}  E > 0 \\
               (1/6) \eta^3,        & \hspace*{1cm}  E = 0 \\
               \eta - \sin(\eta),   & \hspace*{1cm}  E < 0
         \end{array}
             \right.
 \label{eq:phixi}
 \end{equation}
where $t_{B} = t_{B}(r)$ is a third arbitrary function.
The solutions (\ref{eq:R2},\ref{eq:phixi})
have the same evolution as the corresponding {\sc flrw} dust solutions , but
with spatially variable $M$, $E$ and $t_{B}$. In contrast to the {\sc flrw} models,
however, it is quite possible to have all three types of evolution in the
same model \cite{bi:HL85}. In the homogeneous {\sc flrw} case -- $\rho = \rho(t)$ only -- the requirement that
$\eta$  be independent of $r$ at all times in (\ref{eq:R2}, \ref{eq:phixi}) implies that
 \begin{equation}
t_{B} = \mbox{constant},~~~~~~ M \propto {|E|}^{3/2} \,.
 \end{equation}

The function $t_{B}(r)$ is the `bangtime function'. Individual shells of
matter need not all emanate from one single bang event, but originate at
different times as determined by $t_{B}(r)$.  The gradient of $t_B$
generates the decaying modes of the perturbation to an {\sc flrw}
background \cite{bi:S77}.

$M(r)$ is the effective gravitational  mass within $r$.

 The local geometry is determined by $E(r)$, as is evident from its
appearance in the metric, and in fact this function determines the
`embedding angle' \cite{bi:H87}.  Also if we  compare equation
(\ref{eq:motion}) with the Newtonian analogue of a dust cloud we see that
$E$ also acts as an energy potential; that is, locally hyperbolic,
parabolic and elliptic regions occur when $E(r) > 0$, $E(r) = 0$ and $E(r)
< 0$ respectively.  Its gradient generates the growing modes of the
perturbation to an {\sc flrw} background \cite{bi:S77}.

Our method for this investigation is straightforward. We require
that the density be smooth through the origin of our coordinate system.
Thus the spatial gradient of the density has to vanish at $R(t, r = 0)$
for all $t$. This would then impose certain restrictions on the
three arbitrary functions $M(r)$, $E(r)$ and $t_B(r)$ and their derivatives
for this density profile to hold. The change in concavity of the density
profile at the origin is determined by
the sign of the second radial derivative of the density at that point.

For the required density profile we need expressions for the spatial
gradient $R'$ and second and third radial derivatives, $R''$ and $R'''$
respectively, explicitly as a sum of a product of functions of $r$ and
functions of $\eta$.  The full expressions are somewhat
 nasty-looking expressions and are not easily understood without detailed
analysis, so we have merely recorded them in appendix \ref{ap:A}.

 \subsection{Restrictions on the Arbitrary Functions}
 \subsubsection{Shell Crossings }
 \label{sec:ShC}
 We will impose regularity conditions on the spacetime; excluding shell
crossings in particular.
Loosely stated, a shell crossing occurs when an inner spherical shell of
matter moves faster than an outer shell and eventually bursts through. A
locus of points is formed where $ R' = 0 $ and $ R \neq 0 $%
 \footnote{
 Regular maxima in the spatial sections also have $R'=0$, but are not
shell crossings \cite{bi:HL85}.
 }%
 .
 Since the Kretschmann scalar $K = R_{abcd} R^{abcd}$ diverges, one may
consider this to be a `true' singularity%
 \footnote{Other opinions are that these are
 non-physical in the sense that they merely indicate the
impropriety of extending a simplified fluid description too far.
 }%
 .
  In contrast to other studies which utilised the
 high-density regions created by shell crossings as generators of
 large-scale structure, we require the spacetime to be regular and thus
seek to exclude shell crossings.  The necessary and sufficient  conditions
under which shell crossings do not occur were derived by Hellaby and Lake
\cite{bi:HL85}.
 \subsubsection{Behaviour at the Origin}
 \label{sec:origin}

 An origin occurs at $r=0$ when $R(t, r = 0) = 0$ for all $t$.
 On any constant $t$ surface away from the bang or crunch,
we require that \\
 (a) the density $\rho$ be finite, positive, and
 non-zero,
 \begin{equation}
   \frac{M'}{R^2 R'} \rightarrow \kappa \rho_0(t) = \mbox{const} \in
(0,\infty)
 \end{equation}
 (b) the Kretschmann scalar be finite
 \begin{equation}
   K =
   \frac{48 M^2}{R^6} +
   \frac{32 M M'}{R^5 R'} +
   \frac{12 (M')^2}{R^4 (R')^2}
   \rightarrow K_0(t) = \mbox{const} \in (-\infty,\infty)
 \end{equation}
 and (c) the evolution at $r=0$ not be different from its neighbourhood,
so that $(t - t_B)$, $\phi_0(\eta)$ and $\xi(\eta)$ go smoothly to a
finite limit in $(0,\infty)$ as $r \rightarrow 0$.  Equation (\ref{eq:R2})
then gives us the following behaviour of the arbitrary functions near the
origin
 \begin{equation}
   \frac{R}{p} = \frac{R(\pm E)}{M} \rightarrow S_0(t) = \mbox{const} \in
(0,\infty),   \hspace*{1cm}
   q = \frac{M}{(\pm 2 E)^{3/2}} \rightarrow q_0 = \mbox{const} \in
(0,\infty)
 \end{equation}

 If we assume that $E(r)$ and $M(r)$ are analytic at $r=0$, so that they
can be approximated by polynomials in $r$, then we can further deduce
that, as $R \rightarrow 0$,
 \begin{eqnarray}
   E \propto {R}^2 \rightarrow 0, & & ~~~~
   M \propto {R}^3 \rightarrow 0
 \end{eqnarray}
 and similarly
 \begin{equation}
   \dot{R} \propto R \rightarrow 0
 \end{equation}
 Although $M'/M$ \& $E'/E$ both go as $1/R$, the foregoing gives
 \begin{equation}
   \frac{q'}{q} = \left( \frac{M'}{M} - \frac{3 E'}{2E} \right)
\rightarrow \mbox{constant or } 0,   \label{eq:qp0}   \\
 \end{equation}
 Thus we have an {\sc flrw}-like origin%
 \footnote{
 If the density were allowed to approach zero at the origin, other
limiting behaviours of $E$ \& $M$ would be possible.
 }%
 .
 \subsubsection{The Smooth Central Density Criteria}
 The fractional spatial gradient  of the density is obtained by
differentiating  the density with respect to $R$ on a constant
 $t$-slice. Since $R$ is a physically  invariant quantity
 -- the areal radius --  this will give us results which are not
coordinate dependent.  We can take a slice in time in a natural way since
the coordinate time $t$ is also proper time for comoving dust in a
synchronous metric and so also physically invariant.  Furthermore, since
the 3 arbitrary functions $M$, $E$ \& $t_B$ all have physical
interpretations, they are invariant too.

Now the transformation between $(t,r)$ and $(t,R)$ where $R = R(t,r)$
obeys
 \begin{equation}
   \left. \frac{\partial r}{\partial R} \right|_t R' = 1,
   ~~~~~~~~
   \left. \frac{\partial r}{\partial R} \right|_t \dot{R} +
   \left. \frac{\partial r}{\partial t} \right|_R = 0
 \end{equation}
 so for any function of $r$, say $F(r) = F(r(R,t))$, we define
 \[   \left. \frac{\partial F}{\partial R}\right|_t  \equiv
 \partial_{R} \, {F} =
 \frac{dF}{dr} \, \left. \frac{\partial r}{\partial R}\right|_t =
 \frac{F'}{R'} \,.   \]
 So the energy density on a hypersurface of constant time, equation
(\ref{eq:density}), can be written as
 \begin{equation}
 8\pi\rho = \frac{2\, \partial_{R} \,  M}{R^{2}}\,.
 \label{eq:density2}
 \end{equation}
 From the above equation we find the fractional spatial gradient
of the density to be
 \begin{equation}
 \frac{\partial_{R} \,  \rho}{\rho} = \frac{\partial_{RR} \,  M}{\partial_{R} \,  M} -
 \frac{2}{R}
 \label{eq:densgrada}
 \end{equation}
 where we have determined that
 \begin{equation}
 \partial_{RR} \,  M \equiv \left. \frac{{\partial}^2 M}{{\partial R}^2}\right|_t =
(M'' -  (\partial_{R} \,  M) R'') / (R')^2 \,.
 \label{eq:RRM}
 \end{equation}
 We require the density to be finite and its gradient to vanish at the
origin.  Thus, for the required density profile we must have
 \begin{equation}
 \left. \frac{\partial_{R} \,  {\rho}}{\rho}\right|_{r = 0} = -\frac{1}{(R')^2} \hspace*{.1cm}
 \left. \left[2 \hspace*{.1cm} \frac{\left(R'\right)^2}{R} + \hspace*{.1cm} R'' -
 \frac{M''}{M'} \hspace*{.1cm} R' \right]\right|_{r = 0} = 0 \,.
 \label{eq:maxdens}
 \end{equation}

 We find that this gives us four conditions on the arbitrary functions and
their derivatives.  For the details in the hyperbolic case, see appendix
\ref{ap:D}.

 Requiring the density to be $C^1$ at the origin also implies that the
bangtime function $t_B(r)$ must be  at least $C^{1}$ at the origin.  For
the details see appendix \ref{ap:D}.
 \subsection{Evolution of the Second Radial Derivative of the Density}
 To answer the question raised in the introduction, we need to see what
happens to the second radial derivative of the density.  We can obtain an
expression for this quantity by differentiating equation
(\ref{eq:density2}) twice with respect to R on a surface of constant time.
 \begin{equation}
{\frac{\partial_{RR} \,  \rho}{\rho}} =
 \frac{2}{R}\left( \frac{3}{R} -
 \frac{2 \, \partial_{RR} \,  M}{\partial_{R} \,  M} \right) +
 \frac{\partial_{RRR} \, {M}}{\partial_{R} \,  M} \nonumber \\
 \label{eq:rhobb}
 \end{equation}
 where $\partial_{RR} \, M$ is given by (\ref{eq:RRM}) and $\partial_{RRR}
\, M$ is defined as
 \[ \partial_{RRR} \,  M \equiv \left.
 \frac{{\partial}^3M}{\partial{R}^3}\right|_t =
 \frac{1}{(R')^3}(M''' - (\partial_{R} \,  M) R''')
 - \frac{3 (\partial_{RR} \,  M) R''}{(R')^2} \,. \]
 A more explicit form of the above for the hyperbolic case can be obtained
by substituting for $R'$, $R''$ and $R'''$ into equation (\ref{eq:rhobb}).
Again, this is a most
 unpleasant-looking expression and not easily assimilated.  Appendix
\ref{ap:E} contains the result we get after the assumption of a flat
central density has been included.

 \section{Density Profile Inversion: Existence Proof}
 A change in the density contrast will depend on whether factors on the
right hand side of equation (\ref{eq:rhodd2}) change sign with evolution,
or terms of different sign become dominant.  In order to simplify, we have
assumed that all three arbitrary functions have polynomial behaviour at
the origin.  In addition, we imposed the smooth origin condition
(\ref{eq:qp0}), which effectively says that in some neighbourhood of the
origin, the spacetime is tangent to a homogeneous model.  We now single
out the dominant functions of $\eta$ for early and late times in
hyperbolic models, which leads to the following digestible expressions.

 For early times, $\eta \rightarrow 0$,
 \begin{eqnarray}
 \lim_{\eta \rightarrow 0}
 \left.{\frac{\partial_{RR} \, \rho}{\rho}}\right|_{r = 0}
&\simeq&
 \left( \frac{12E}{M'} \frac{1}{\eta} \right)^{3} 6 \sqrt{2E} t_{B}'
 \left[
\frac13
 \left(
 \frac{t_{B}'''}{t_{B}'} - \frac{M'''}{M'}
 \right)
 \right.
 \nonumber \\
&+&
 \left.
 \left(
  \frac{M''}{M'}
- \frac{t_{B}''}{t_{B}'}
 \right)
 \left(\frac{M''}{M'}
- \frac{11}{9} \frac{M'}{M}
 \right)
 + \frac{4}{9} \frac{{M'}^2}{M^2}
\right] \,;
 \label{eq:rhodd_a}
 \end{eqnarray}
with $R'$ at early times  given by
 \[
 \left. R'\right|_{\eta \rightarrow 0} \simeq \frac{1}{12} \frac{M'}{E} \eta \,.
 \]

 For late times we note that $\sinh \eta \rightarrow \cosh \eta$, $\cosh
\eta - 1 \rightarrow \cosh \eta$ where $\cosh \eta \rightarrow
e^{\eta}\,/\,2$ as $\eta \rightarrow \infty$.  Therefore for large $\eta$,
and at the origin,
 \begin{eqnarray}
 \lim_{\eta \rightarrow \infty}
 \left.{\frac{\partial_{RR} \,\rho}{\rho}}\right|_{r=0}
&\simeq&
 \left(\frac{8E^{3}}{M E'^{2}} \, \frac{1}{\cosh \eta} \right)^{2}
 \left[\frac{1}{4} \left(\frac{M'''}{M'} - \frac{E'''}{E'} \right)
 \frac{{E'}^{2}}{E^2}
 - \frac12
 \left(\frac{E'}{E} \frac{M''}{M'} -  \frac{E''}{E} \right) \, \frac{M''}{M}
\right.
 \nonumber \\
 &-&
 \left.
  \left(  \frac12
 \frac{{E'}^{2}}{E^{2}} - \frac79 \frac{E'}{E} \frac{M'}{M}
  + \frac19\frac{{M'}^{2}}{M^{2}} \right) \, \frac{M''}{M}
  - \frac{5}{18} \frac{{M'}^{2}}{M^2} \frac{E''}{E}
\right.
 \nonumber \\
 &+&
 \left.
\frac13  \frac{{M'}^{2}}{M^2}\left(
  \frac{19}{12} \frac{{E'}^{2}}{E^2} - \frac{16}{9} \frac{M'}{M} \frac{E'}{E}
  + \frac{4}{9} \frac{{M'}^{2}}{M^{2}}
  \right)
  \right] \,,
 \label{eq:rhodd_b}
 \end{eqnarray}
where $R'$ at late times is given by
 \[
 \left. R' \right|_{\eta \rightarrow \infty} \simeq \frac{1}{4} \frac{M
E'}{E^2} \cosh \eta \,.
 \]

 To what extent can we tailor the evolution of $\rho$ by choosing
the 3 arbitrary {\sc ltb} functions?  We recall that there are four
restrictions on the nine quantities $E$, $E'$, $E''$, $M$, $M'$, $M''$,
$t_B$, $t_B'$  and $t_B''$ given by equations
 (\ref{eq:cond1})-(\ref{eq:cond4}) ensuring a flat central density.  This
leaves us with the freedom to fix five of the above at $r=0$.  In addition
there are also the conditions for a regular origin (section
\ref{sec:origin}) and those for no shell crossings \cite{bi:HL85}.
However, the former conditions do not provide any additional choice
restrictions once we have specified a flat density at the origin, and the
latter are inequality constraints which only limit the range of choice, so
are not as severe as the others.

 We see in the early time expression (\ref{eq:rhodd_a}) there appear terms
involving the three derivatives of $t_B$ which do not occur in the late
time expression (\ref{eq:rhodd_b}).  This allows us to fix the early time
behaviour of the density.  And, likewise, the three derivatives of $E$
occur at late times but not at early times for the relative change in
concavity of the density.  Thus, in principle, we should be able to
independently fix the late time behaviour as well due to this freedom.  In
effect, we have sufficient freedom to model the density as being overdense
initially, and subsequently evolving to an underdense state, or indeed,
vice versa.  Moreover, it is conceivable that the middle time behaviour
could be separately specified, as there are still the derivatives of $M$
to play with.
 \section{Specific Models}
 We will consider an initial overdensity changing to an underdensity. So,
at early times at the origin, we want the concavity to be negative and at
late times, positive.  We consider `exact perturbations'%
 \footnote{
 This is just a mathematical device.  No averaging or matching procedure
to define a background {\sc flrw} model has been employed
 --- there is no `gauge problem' in the sense of cosmological
perturbations relative to a background.
 }
 of an {\sc flrw} model in the following way:
 \begin{eqnarray}
   M(r) &=& M_0 r^3 (1 + \alpha(r)), ~~~\alpha(0) = 0 \,; \\
   2\,E(r) &=&  r^2 (1 + \beta(r)), ~~~\beta(0) = 0 \,; \\
   t_B(r) &=& \gamma(r), ~~~\gamma(0) = 0 \,.
 \end{eqnarray}
 These ensure the origin conditions of section \ref{sec:origin} are
satisfied. However, they do not necessarily satisfy the restrictions
imposed by
 (\ref{eq:cond1})-(\ref{eq:cond4}).

 We are also preventing
 shell-crossing singularities from interfering.  These conditions are, for
$t > t_B$, $R' > 0$ and hyperbolic models $ER/M > 0$,
 \[
   t_B' \leq 0\,,~~ E' > 0~~\mbox{and}~~ M' \geq 0 \,.
 \]
 For perturbation functions of the form
 \begin{eqnarray}
   \alpha(r) &=& Ar^a\,, \\
   \beta(r) &=& Br^b\,, \\
   t_B(r) &=& Cr^c\,,
 \end{eqnarray}
 the requirement of no
 shell-crossings leads to the following restrictions on the constants $A$,
$B$ and $C$:
 \begin{eqnarray}
t_B' \leq 0 &\Rightarrow& C \leq 0  \\
M' \geq 0   &\Rightarrow& A \geq -\frac{3}{(3 + a) \, r^a}  \\
E' > 0      &\Rightarrow& B > -\frac{2}{(2 + b) \, r^b} \,.
 \label{eq:sc}
 \end{eqnarray}
 The smooth central density criteria
 --- that is, in this case, (\ref{eq:cond2}) and  (\ref{eq:cond4}) for
$t_B(r)$ or (\ref{eq:cond1}) and (\ref{eq:cond3})  for $M(r)$ and $E(r)$
 --- impose the following restrictions on $A$, $B$ and $C$ (Tables
\ref{ta:a_b} and \ref{ta:c}).  For simplicity, we will only investigate
models where $a$, $b$ and $c$ are (positive) natural numbers.
 \begin{center}
 ----------------- \\
 Table 1 goes here \\
 ----------------- \\[6mm]
 ----------------- \\
 Table 2 goes here \\
 -----------------
 \end{center}

 Choosing a bangtime function $t_B = Cr^2$, substitution into equation
(\ref{eq:rhodd_a}) shows that the relative concavity of the density at
early times can be fixed as negative by choosing $C$ negative, making
$t_B$ a decreasing function; in fact,
 \[
   \lim_{\eta \rightarrow 0} \left. \frac{\partial_{RR} \, \rho}{\rho}
   \right|_{r=0} = \frac{160 C}{M_0^3} \,.
 \]
 This automatically satisfies the first requirement for no shell crossings
to occur as well.  A choice of $t_B$ which is of higher power gives
$\displaystyle{\frac{\partial_{RR} \, \rho}{\rho} = 0}$ at the origin at
early times.  Choosing $t_B$ as a linear function results in a vanishing
bangtime perturbation as can be seen from Table {\ref{ta:c}.

 We use equation (\ref{eq:rhodd_b}) to determine the late time behaviour.
The results after use of Table \ref{ta:a_b} are tabulated below (Table
\ref{ta:dpimod}).
 \begin{center}
 ----------------- \\
 Table 3 goes here \\
 -----------------
 \end{center}

 Clearly there are a wide variety of models which can change concavity at
the origin and which also have no
 shell-crossing singularities.

 We will illustrate the phenomenon on a model which has quadratic
perturbation functions
 --- that is; $a$, $b$ and $c$ are all equal to two.  We choose $A = 1
\times 10^{2}$, $B = 1 \times 10^{-6}$ and $C = -3 \times 10^{-8}$.  Since
we are only interested in qualitative results, we may put $M_0 = 1$.

 The density profile this specifies is plotted for a sequence of cosmic
time ($t$) values in figures 1-4.  The units (cosmological time, length,
mass and density units) are converted as follows
 \\[2mm]
 ${}$   \hfill
 \begin{tabular}{lll}
  1~ctu         &=& 2.005 $\times 10^9$~yrs \\
  1~clu         &=& 6.146 $\times 10^8$~pc \\
  1~cmu         &=& 1.285 $\times 10^{22}$~$M_{\odot}$ \\
  1~cmu/clu$^3$ &=& 3.746 $\times 10^{-27}$~g/cc~.
 \end{tabular}
 \hfill   ${}$

 \begin{center}
 --------------------- \\
 Figures 1 - 4 go here \\
 ---------------------
 \end{center}

 \section{Implications and Discussion}

 We have shown that, for the simplest inhomogeneous cosmologies
 --- the LTB models for spherically symmetric dust
 --- a change from central density maximum to central density minimum (or
vice-versa) during the evolution of the inhomogeneity is entirely
possible, and a numerical example was presented.  Indeed, given that the
early and late time limits of the concavity of the central density profile
depend on separate arbitrary functions that have no necessary connection,
it would be surprising if profile inversions were not common.  The models
considered are completely physically reasonable for
 post-decoupling inhomogeneities.

 Perhaps the most important implications of this work derive from the
existence proof of the possibility of density profile inversion and
density waves%
 \footnote{
 as has been numerically discovered previously in many studies (in {\sc
ltb} and related models) on large scale structures, mentioned in the
introduction.
 }
 in the {\sc ltb} model.  In the real universe, which is much more complex
than this model, we expect waves to be generic \cite{bi:EHM90}.  The
crucial element in our investigation is the importance of the bangtime
function $t_B$ and its derivatives at early times.  We may recall from
section \ref{sec:descr} that $t_B'$ generates the decaying mode and $E'$
the growing mode to {\sc rw} perturbations.  The overdensity occurs at
early times because we choose $t_B$ in such a way that the result is an
overdensity and, in a similar fashion, the underdensity occurs at a late
time because we choose $E$ such that it gives that particular type of
density profile.  Perhaps the reason why the effect obtained here has not
been discussed before is because most studies consider linearised
perturbations which have the ultimate effect of neglecting the decaying
mode.

 The most common model of structure formation assumes Cold Dark Matter
({\sc cdm}) with a
 Harrison-Zel'dovich spectrum of initial perturbations.  Observations
indicate that {\sc cdm} predictions on large scales and small scales are
incompatible.  In particular, standard {\sc cdm} has trouble reproducing
the large velocity dispersion of luminous matter from the stationary
standpoint.  Realising the fairly universal failings of standard structure
formation theories to explain bulk flow statistics, one might argue that
there is some fundamental assumption that must be
 re-evaluated unless there is a radically different  process responsible
for structure in the universe.  It seems natural to ask if our results
might go at least some way in solving these problems.

 In the linear theory of structure formation, the topology of density
contour surfaces does not change.  No links are formed and no chains are
broken --- the genus of the surface is unchanged since the process is
continuous.  When nonlinearity is important, the genus of the surfaces
evolves as clumps and bubbles form.  Even though the statistics today may
be
 non-Gaussian, their structure today will vary depending on the
Gaussianity of the initial distribution.  However, the way that this
occurs will be different if the density profile inverts and if density
waves are present, since no longer will overdense regions simply grow
monotonically.  There will be an interaction of spatial and temporal
density gradients.  This means that density waves must be included if a
correct interpretation of topological studies of structure formation is to
be obtained.  The change to the topology of the constant density contours,
comparing density waves and
 no-density waves scenarios, should be examined.

 It should be clarified that the density waves discussed here are due to
motion of the density maximum through the comoving frame, so that a galaxy
that is in the density peak at one time may be outside it at a later or
earlier time.  This effect may be superimposed on the galaxy flow.

 A direct effect of this work is its implications for the transfer
function used ubiquitously in standard structure formation theories
whereby luminous matter congregates in the peaks of the underlying mass
distribution%
 \footnote{
 Shear may alter this but this is not well established yet.
 }%
 .
 These peaks do not move; in the sense that they remain attached to the
same world line as time evolves.  The only change that happens is the
infall of matter about these peaks so that the density contrast increases.
There is spatial flow of matter, but the spatial distribution of extrema
of the initial density field remains invariant.  This invariance is broken
when density profile inversions occur.  The effect is to (amongst other
things) change the form of the transfer function.  We could reasonably
speculate that the transfer function becoming more complicated may perhaps
allow one to take a standard scale invariant spectrum and fit it to small,
large and intermediate constraints.

 Another way of viewing this is that we may not be able to rely on
luminous matter being an accurate tracer of total cosmic density,
since the density peaks that triggered galaxy formation may have moved on.
Similarly, the velocity imparted to forming galaxies may no longer be that
of the unseen matter component.  The result of the gravitational
interaction of the luminous and dark components may be observed flows
towards regions which do not seem to be density concentrations.  However,
such
 two-component effects are beyond the present study.

 \vspace*{3mm}

 \noindent{\Large \bf Acknowledgements}~~~NM is pleased to thank Bruce
A.C.C. Bassett for stimulating discussions on this and other related
work.\\
 The computer algebra package Maple was used to check many of the
equations obtained at various stages of development. \\
 CH thanks the NRF for a research grant.

 

 \appendix

 \newpage
 \section{Spatial Derivatives of Areal Radius}
 \label{ap:A}

We find
 \[\frac{\partial{R}}{\partial{r}} =
 \frac{p}{2} u \phi_0 + \frac{p}{2} \frac{d\phi_0}{d\eta}\, 1/\, \left(
 \frac{d\xi}{d\eta} \right) \frac{\partial{\xi}}{\partial{r}}
 \]
 where $\phi_0$ and $\xi$ are given by (\ref{eq:phixi}), and $u(r)$ has
been defined as
 \begin{equation}
   u \equiv (\ln p)' =  \frac{M'}{M} - \frac{E'}{E}\,.
     \label{eq:u}
 \end{equation}
 After some manipulation we obtain
 \begin{equation}
   \frac{\partial{R}}{\partial{r}} =
   \frac{p}{2} v \phi_2 - \frac{p}{q}\,{t_B}' \phi_1 +
   \frac{p}{2} u \phi_0
     \label{eq:Rd}
 \end{equation}
 where $\phi_1$ to $\phi_9$ are given in (\ref{eq:phi1})-(\ref{eq:phi9}),
and $v(r)$ is defined by
 \begin{equation}
   v \equiv \left( \ln \frac{1}{q} \right)' =
   \frac{3E'}{2E} - \frac{M'}{M} \,.
 \label{eq:v}
 \end{equation}
 We proceed in a similar fashion to obtain an expression for the second
radial derivative:
 \begin{eqnarray}
 \frac{\partial^2{R}}{\partial{r}^2} &=&  \frac{p}{2} v^2 \phi_5 - \frac{p}{q}\, {t_B}' v
 \phi_4 +  \frac{2\,p\,({t_B}')^2}{q^2} \phi_3  + \frac{p}{2}\,(v' + 2uv) \phi_2
 \nonumber \\
                          &-& \frac{p}{q}\, {t_B}'\,\left( w + u \right)
 \phi_1  + \frac{p}{2}\, (u' + u^{2}) \phi_0
 \label{eq:Rddb}
 \end{eqnarray}
where
 \begin{equation}
w(r) \equiv \left( \ln \frac{p}{q}\,{t_B}' \right)' = \frac{E'}{2E} +
 \frac{{t_B}''}{{t_B}'}\,.
 \label{eq:w}
 \end{equation}
 And for the third derivative
 \begin{eqnarray}
 \frac{\partial^3{R}}{\partial{r}^3} &=& \frac{p}{2}v^3\phi_{9} - \frac{p}{q}\,
{t_B}'\,v^2\phi_{8} +  \frac{2\,p \,({t_B}')^2}{q^2}v\phi_{7} -
4p \,\left( \frac{{t_B}'}{q} \right)^3 \phi_{6} \nonumber \\
                          &+& \frac{3}{2}\,pv\,(uv + v')\phi_{5} -
 \frac{3p}{2q}\, {t_B}'\,(v' + uv + wv)\phi_{4} \nonumber \\
                          &+&  \frac{6\,p\,({t_B}')^2}{q^2} \, w \,\phi_{3}
+ \frac{p}{2}(3u^2 v + 3uv' + 3u'v + v'')\phi_{2} \nonumber \\
                          &-& \frac{p}{2q}\, {t_B}'\,(v'+ 2w' + 4u' + 2w^2 +
2uw + 2u^2 + uv - wv) \phi_{1}  \nonumber \\
                          &+& \frac{p}{2}(u^3 +  3uu' + u'') \phi_{0} \,.
 \end{eqnarray}
 The above derivatives of $R$ have been expressed in terms of $u$, $v$ and
$w$ because if written in terms of $M$, $E$ and $t_{B}$ the expressions
become a bit messy and are not very useful in that form at this stage.
Quantities determined later will be expressed in terms of the latter
variables when appropriate.

 The various functions of $\eta$ used above are
 \begin{eqnarray}
 \phi_1(\eta) &=& {\frac {\sinh \eta}{\phi_0}}   \label{eq:phi1}   \\
 \phi_2(\eta) &=& {\sinh \eta} \,\frac{\xi}{\phi_0}\\
 \phi_3(\eta) &=& - \left( \frac{1}{\phi_0} \right)^2\\
 \phi_4(\eta) &=& \phi_1 - 2 \frac{\xi}{\phi_0^{2}}\\
 \phi_5(\eta) &=& \phi_2 - \frac{\xi^2}{\phi_0^{2}}\\
 \phi_6(\eta) &=& 2 \frac{\sinh \eta}{{\phi_0}^4}      \\
 \phi_7(\eta) &=& 3 \phi_3 + 6 \sinh \eta \frac{\xi}{{\phi_0}^4}      \\
 \phi_8(\eta) &=& \phi_1 - \frac{6 \xi}{{\phi_0}^2}  + 6 \sinh \eta
 \frac{\xi^2}{{\phi_0}^4}      \\
 \phi_9(\eta) &=& \phi_5 + 2 \sinh \eta \frac{\xi^3}{{\phi_0}^4} - \frac{2
 \xi^2}{{\phi_0}^2} = \phi_2 - \frac{3 \xi^2}{{\phi_0}^2} + 2 \sinh \eta
 \frac{\xi^3}{{\phi_0}^4}   \label{eq:phi9}
 \end{eqnarray}

 \section{The smooth central density criteria}
 \label{ap:D}
 We want the density at the origin to be flat at all times.

 We substitute $R'$ and $R''$ into  equation (\ref{eq:maxdens}) to obtain
restrictions on the arbitrary functions $E(r)$, $M(r)$ and $t_B(r)$ for
(\ref{eq:maxdens}) to hold.

 We find that
 \begin{eqnarray}
&& \frac{1}{(R')^2} \hspace*{.1cm} \left[ \frac{p}{2}v^2 \left( \phi_5 + 2
 \frac{{\phi_2}^2}{\phi_0} \right) -
 \frac{p\,t_{B}'}{q} \,v\, \left( \phi_4 + 4\frac{\phi_1 \phi_2}{\phi_0}
 \right) + \right.   \nonumber \\
&&
 \left. \frac{2\,p\,(t_{B}')^2}{q^2} \left( \phi_3 +
 2\frac{{\phi_1}^2}{\phi_0} \right) - \frac{p}{2} \left( \frac{M''}{M'}v -
 6uv - v' \right) \phi_2 - \right.    \nonumber \\
&& \left.
 \frac{p\,t_{B}'}{q} \left( w + 5u - \frac{M''}{M'} \right) \phi_1 -
 \frac{p}{2} \left( \frac{M''}{M'}u - 3u^2 - u' \right) \phi_0 \right]
 \label{eq:max}
 \end{eqnarray}
 must vanish at the origin.
 Here the $\phi_{i}$ are all functions of $\eta$ and are defined along
with $u(r)$, $v(r)$ and $w(r)$ in appendix \ref{ap:A}.  Since the
functions of parameter time $\eta$ are linearly independent of each other,
it follows that each of the terms in equation (\ref{eq:max}) must vanish
separately.

From the first term we get
 \begin{equation}
 \left. \frac{1}{(R')^2} {p} \, v^2 \right|_{r=0} = 0
 \label{eq:cond1}
 \end{equation}
whilst the third gives
 \begin{equation}
 \left. \frac{1}{(R')^2} p \left( \frac{t_{B}'}{q} \right)^2
 \right|_{r=0} = 0 \,.
 \label{eq:cond2}
 \end{equation}
 Note that the constraint arising from the second term is satisfied if the
first (\ref{eq:cond1}) and third (\ref{eq:cond2})  constraints are.  The
fourth term combined with equation (\ref{eq:cond1}) expands to
 \begin{equation}
 \left. \frac{1}{(R')^2} p {\cal A}
 \right|_{r=0} = 0 \, , ~~~~~~~~
 {\cal A} \equiv \frac{E'M''}{EM'} - \frac{1}{2}\left(\frac{E'}{E}\right)^2
 - \frac{E''}{E}
 \label{eq:cond3}
 \end{equation}
 The fifth term combined with the second in equation (\ref{eq:max}) yields
 \begin{equation}
 \left. \frac{1}{(R')^2} p \frac{t_{B}'}{q} {\cal B}
 \right|_{r=0} = 0 \, , ~~~~~~~~
 {\cal B} \equiv \frac{M''}{M'} - 2\frac{M'}{M} - \frac{{t_B}''}{{t_B}'}\,.
 \label{eq:cond4}
 \end{equation}
 The last term produces a condition  equivalent to equation (\ref{eq:cond1})
combined with (\ref{eq:cond3}).

 We show now that the requirement of having the density smooth through the
origin, leading to the constraints in appendix \ref{ap:D}, implies that
the bangtime function $t_B(r)$ must have vanishing first spatial
derivative at the origin.  We prove this using the coordinate choice $R'
\sim 1 \Rightarrow R \sim r$.  As before, we can take the origin to be at
$r = 0$ without loss of generality.  In these coordinates, the assumption
of analytic arbitrary functions near $r=0$ gives $p = M/(\pm E) \sim r^1$,
$q = M/(\pm E)^(3/2) \sim r^0$, since $M \sim r^3$ and $E \sim r^2$.  The
relation which is of importance to us here is equation (\ref{eq:cond4}).
It says that
 \begin{equation}
   \left. \frac{1}{(R')^2} p \frac{t_{B}'}{q} \left( \frac{M''}{M'} -
   2\frac{M'}{M} - \frac{{t_B}''}{{t_B}'} \right) \right|_{r=0} = 0 \,.
         \label{eq:flatbang}
 \end{equation}
 In this expression, we must have
 \begin{equation}
 \left(
 \frac{M''}{M'} - 2 \frac{M'}{M}
 \right)
 \sim \frac{-4}{r}
 \label{eq:term1}
 \end{equation}
 Therefore, either $t'_B \sim r^{-4}$ to make ${\cal B}$ zero, or ${\cal
B} = (M''/M' - 2M'/M - t''_B/t'_B)$ diverges as $1/r$ or faster.  In the
former case we get $t_B \sim r^{-3}$, which is not reasonable
 --- either the universe is infinitely old at the origin only, or it will
not emerge from the bang for an infinite time.  In the latter case
$p(M''/M' - 2M'/M)$ is constant, so
we require
 \begin{equation}
   t_{B}' p \left( \frac{M''}{M'} - 2\frac{M'}{M} \right) \rightarrow 0
   ~~~~~~\mbox{and}~~~~~~
   p {t_B}'' \rightarrow 0
 \end{equation}
 which, near $r=0$, implies
 \begin{equation}
   t_B \sim r^c~,~~~~ c > 1 \, .
 \end{equation}

 \section{The Relative Concavity of the Density}
 \label{ap:E}

 With repeated application of equations
 (\ref{eq:cond1})-(\ref{eq:cond4}) ensuring a smooth central density (in
particular, taking the bangtime derivative to be vanishing at the origin);
and using the variables defined by equations (\ref{eq:p}), (\ref{eq:q}),
(\ref{eq:v}), (\ref{eq:cond3}) and (\ref{eq:cond4}), we find a greatly
expanded form of equation (\ref{eq:rhobb}).

 \begin{eqnarray}
 \lefteqn{
 \left.
 \frac{\partial_{RR} \, \rho}{\rho}  \times \left( R' \right)^4
 \right|_{r = 0}
 }
 \nonumber \\
 &=&
 \left[ \frac{4\,p^2 {v}^{3}}{3} \frac{M'}{M} \right]
 \left(\phi_{9} \phi_{0} +  \phi_{5} \phi_{2}
 -  \frac{8\phi_{2}^{3}}{\phi_{0}}
 \right)
 \nonumber \\
 &+&
 {p^2 \over 4}
 \left[
 {v}^{2}
 \left(
 {\frac {{\it M'''}}{{\it M'}}}-{\frac {{\it E'''}}{{\it E'}}}
 \right)
 + {v}\frac{{\it M'}}{M}\,
 \left(
 {\frac {{\it M'''}}{{\it M'}}}-{\frac {{\it E'''}}{{\it E'}}}
 \right )
 \right.
 \nonumber \\
 &&
 \left.
 + \frac{ 13\,M'\,{\cal A} {v} }{ 2\,M }
 - \frac{ 9\,M''{\cal A} {v} }{ 2\, M'}
 -{\frac {{\it M''}\,{v}^{3}}{{\it M'}}}
 -{\frac {2\,{\it M''}\, {v}^{2}}{M}}
 \right.
 \nonumber \\
 &&
 \left.
 -{\frac {{\it M''}\,{\it M'}\,{v}}{M^{2}}}
 -{\frac {29\,M'{v}^{3}}{9\,M}}
 +{\frac {26\,{\it M'}^{2}{v}^{2}}{9\,M^{2}}}
 +{\frac {4\,{\it M'}^{3}{v}}{9\,M^{3}}}
 \right]
 \phi_{2}^{2}
 \nonumber\\
 &+&
 {\frac{2\,p^2 t_B'}{q}}
 \left[
 -{v}
 \left(
 {\frac {{\it M'''}}{{\it M'}}}-{\frac {{\it t_{B}'''}}{{\it t_{B}'}}}
 \right)
 -{v}
 \left(
 {\frac {{\it M'''}}{{\it M'}}}-{\frac {{\it E'''}}{{\it E'}}}
 \right)
 -\frac{{\it M'}}{M} \,
 \left(
 {\frac {{\it M'''}}{{\it M'}}}-{\frac {{\it E'''}}{{\it E'}}}
 \right)
 \right.
 \nonumber \\
 &&
 \left.
 +{\frac {3\,{\it M''}\,{\cal B}{v}}{{\it M'}}}
 -{\frac {19\,{\it M'}\,{\cal B}{v}}{3\,M}}
 +{\frac {8\,{\it M''}\,{v}}{M}}
 +{\frac {{\it M''}\,{\it M'}}{M^{2}}}
 -{\frac {94\,{\it M'}^{2}{v}}{9\,M^{2}}}
 -{\frac {4\,{\it M'}^{3}}{9\,M^{3}}}
 \right]
 \phi_{1} \phi_{2}
 \nonumber \\
 &+&
 {p^2 \over 4}
 \left[
 {\frac {3\,{\it M'}\,{\cal A}{v}}{2\,M}}
 +{\frac {17\,M'{v}^{3}}{9\,M}}
 +{\frac {7\,{\it M'}^{2}{v}^{2}}{9\,M^{2}}}
 \right]
 \phi_{5} \phi_{0}
 - 2\, { \frac{p^2 t_B'}{q} }
 \left[
 {\frac {7\,{\it M'}^{2}{v}}{9\,M^{2}}}
 \right]
 \phi_{4} \phi_{0}
 \nonumber \\
 &+&
 {p^2 \over 4}
 \left[
 {4 \over 3}\,
 \left( {v}+{\frac {{\it M'}}{M}} \right)
 \left( {\frac {{\it M'}}{4\,M}}-{v} \right)
 \left(
 {\frac {{\it M'''}}{{\it M'}}}-{\frac {{\it E'''}}{{\it E'}}}
 \right)
 \right.
 \nonumber \\
 &&
 \left.
 -{\frac {3\,{\it M''}\,{\cal A}}{2\,M}}
 +{\frac {5\,{\it M'}^{2}{\cal A}}{6\,M^{2}}}
 +{\frac {6\,{\it M''}{\cal A}{v}}{{\it M'}}}
 -{\frac {35\,{\it M'}{\cal A}{v}}{3\,M}}
 \right.
 \nonumber \\
 &&
 \left.
 +{\frac {4\,{\it M''}\,{v}^{3}}{3\,{\it M'}}}
 +{\frac {7\,{\it M''}\,{v}^{2}}{3\,M}}
 +{\frac {2\,{\it M''}\,{\it M'}\,{v}}{3\,M^{2}}}
 -{\frac {{\it M'}^{2}{\it M''}}{3\,M^{3}}}
 \right.
 \nonumber \\
 &&
 \left.
 -{\frac {38\,M'{v}^{3}}{27\,M}}
 -{\frac {47\,{\it M'}^{2}{v}^{2}}{9\,M^{2}}}
 -{\frac {4\,{\it M'}^{3}{v}}{27\,M^{3}}}
 +{\frac {4\,{\it M'}^{4}}{27\,M^{4}}}
 \right]
 \phi_{2} \phi_{0}
 \nonumber \\
 &+&
 { \frac{2\,p^2 t_B'}{3\,q} }
 \left[
 \left( 2\,{v}-{\frac {{\it M'}}{M}} \right)
 \left(
 {\frac {{\it M'''}}{{\it M'}}}-{\frac {{\it t_{B}'''}}{{\it t_{B}'}}}
 \right)
 +2\, \left( {v}+{\frac {{\it M'}}{M}} \right)
 \left(
 {\frac {{\it M'''}}{{\it M'}}}-{\frac {{\it E'''}}{{\it E'}}}
 \right)
 \right.
 \nonumber \\
 &&
 \left.
 +{\frac {3\,{\it M''}\,{\cal B}}{M}}
 -{\frac {11\,{\it M'}^{2}{\cal B}}{3\,M^{2}}}
 -{\frac {6\,{\it M''}\,{\cal B}{v}}{{\it M'}}}
 +{\frac {79\,{\it M'}\,{\cal B}{v}}{6\,M}}
 -{\frac {16\,{\it M''}\,{v}}{M}}
 \right.
 \nonumber \\
 &&
 \left.
 +{\frac {4\,{\it M''}\,{\it M'}}{M^{2}}}
 +{\frac {65\,{\it M'}^{2}{v}}{3\,M^{2}}}
 -{\frac {46\,{\it M'}^{3}}{9\,M^{3}}}
 \right]
 \phi_{1} \phi_{0}
 \nonumber \\
 &+&
 {p^2 \over 4}
 \left[
 {2 \over 9}\,
 \left( {v}+{\frac {{\it M'}}{M}} \right)
 \left( {{2\,{v}}}-{\frac {{\it M'}}{M}} \right )
 \left(
 {\frac {{\it M'''}}{{\it M'}}}-{\frac {{\it E'''}}{{\it E'}}}
 \right )
 \right.
 \nonumber \\
 &&
 \left.
 +{\frac {{\it M''}\,{\cal A}}{M}}
 -{\frac {5\,{\it M'}^{2}{{\cal A}}}{9\,M^{2}}}
 -{\frac {2\,{\it M''}\,{\cal A} {v} }{{\it M'}}}
 +{\frac {38\,{\it M'}\,{{\cal A}}{{v}}}{9\,M}}
 \right.
 \nonumber \\
 &&
 \left.
 -{\frac {4\,{\it M''}\,{v}^{3}}{9\,{\it M'}}}
 -{\frac {2\,{\it M''}\,{v}^{2}}{3\,M}}
 +{\frac {2\,{\it M'}^{2}{\it M''}}{9\,M^{3}}}
 +{\frac {76\,M'{v}^{3}}{81\,M}}
 \right.
 \nonumber \\
 &&
 \left.
 +{\frac {50\,{\it M'}^{2}{v}^{2}}{27\,M^{2}}}
 -{\frac {8\,{\it M'}^{3}{v}}{81\,M^{3}}}
 -{\frac {8\,{\it M'}^{4}}{81\,M^{4}}}
 \right]
 \phi_{0}^{2}
 \label{eq:rhodd2}
 \end{eqnarray}

 \newpage
 \section{Tables}

 \begin{table}[h]
 \begin{center}
 \begin{tabular}{|l|l|}
 \hline
 $c = 1$      & $C = 0$        \\
 \hline
 $c \geq 2$   & no restrictions \\
 \hline
 \end{tabular}
 \caption{
 \label{ta:c}
 Restrictions imposed on the perturbation $t_B$ by the requirement of a
$C^1$ central density ($\forall \; c \in {\N}^{+}$).
 }
 \end{center}
 \end{table}

 \begin{table}[h]
 \begin{center}
 \begin{tabular}{|l|l|l|}
 \hline
              & $a = 1$              & $a \geq 2$      \\
 \hline
 $b = 1$      & $A = B$              & $B = 0$         \\
 \hline
 $b \geq 2$   & $A = 0$              & no restrictions \\
 \hline
 \end{tabular}
 \caption{
 \label{ta:a_b}
 Restrictions imposed on the perturbations to $E$ and $M$ by the
requirement of a flat central density ($\forall a, b \in {\N}^{+}$).
 }
 \end{center}
 \end{table}

 \begin{table}[h]
 \begin{center}
 \begin{tabular}{|l|l|l|l|}
 \hline
              & $a = 1$ & $a = 2$ & $a \geq 3$ \\
 \hline
 $b = 1$ &     $1/2\,(-A/r + A^2)$
                        & $2A$
                                  & $0$ \\
 \hline
 $b = 2$       & $-3B$
                        & $2A - 3B$
                                  & $-3B$ \\
 \hline
 $b \geq 3$    & $0$
                        & $2A$
                                  & $0$ \\
 \hline
 \end{tabular}
 \caption{
 \label{ta:dpimod}
 The sign of ${({\partial_{RR} \,\rho})\,/\,{\rho}}$ at the origin at late
times.
 }
 \end{center}
 \end{table}

 \newpage
 \section{Figures}

 \begin{figure}[h]
 \epsffile[0 0 375 266]{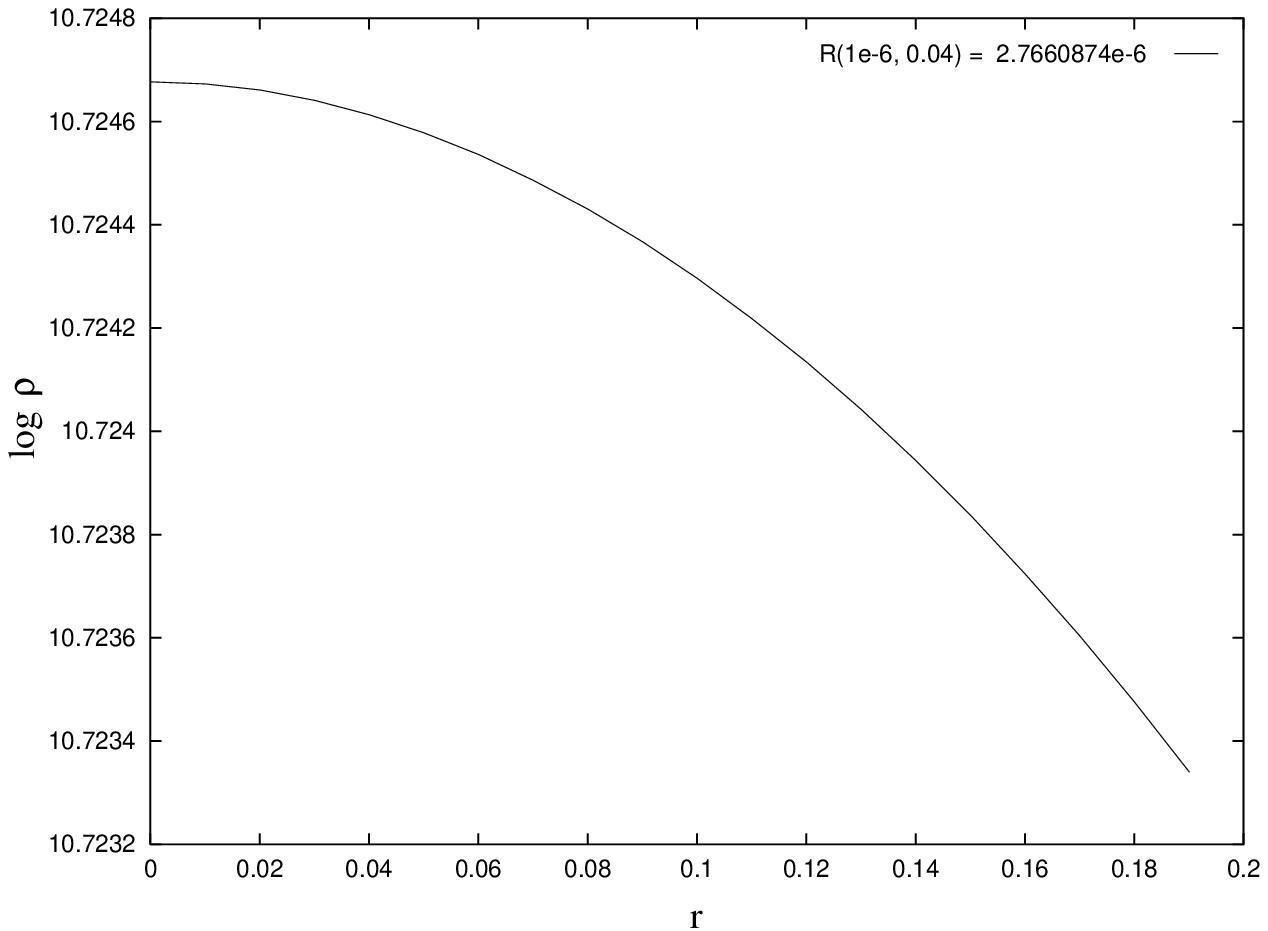}
 \caption{
 The density profile on a worldline at an early time ($t = 1 \times 10^{-
6}$~ctu $\approx 2000$~yrs).  In these figures, the core is taken to be
the value of $R$ at the comoving radius $r=0.04$ and they all use base 10
logs.  At this time the value corresponds to an overdensity of about
$1.7$~kpc in diameter which corresponds to the size of a small galaxy
today.  The units in all the figures are cosmological.  The value of the
density at the origin is $\rho_0 \approx 2.0 \times 10^{-16}$~g/cc.
 }
 \label{fi:dpi1}
 \end{figure}

 \begin{figure}[h]
 \epsffile[0 0 389 265]{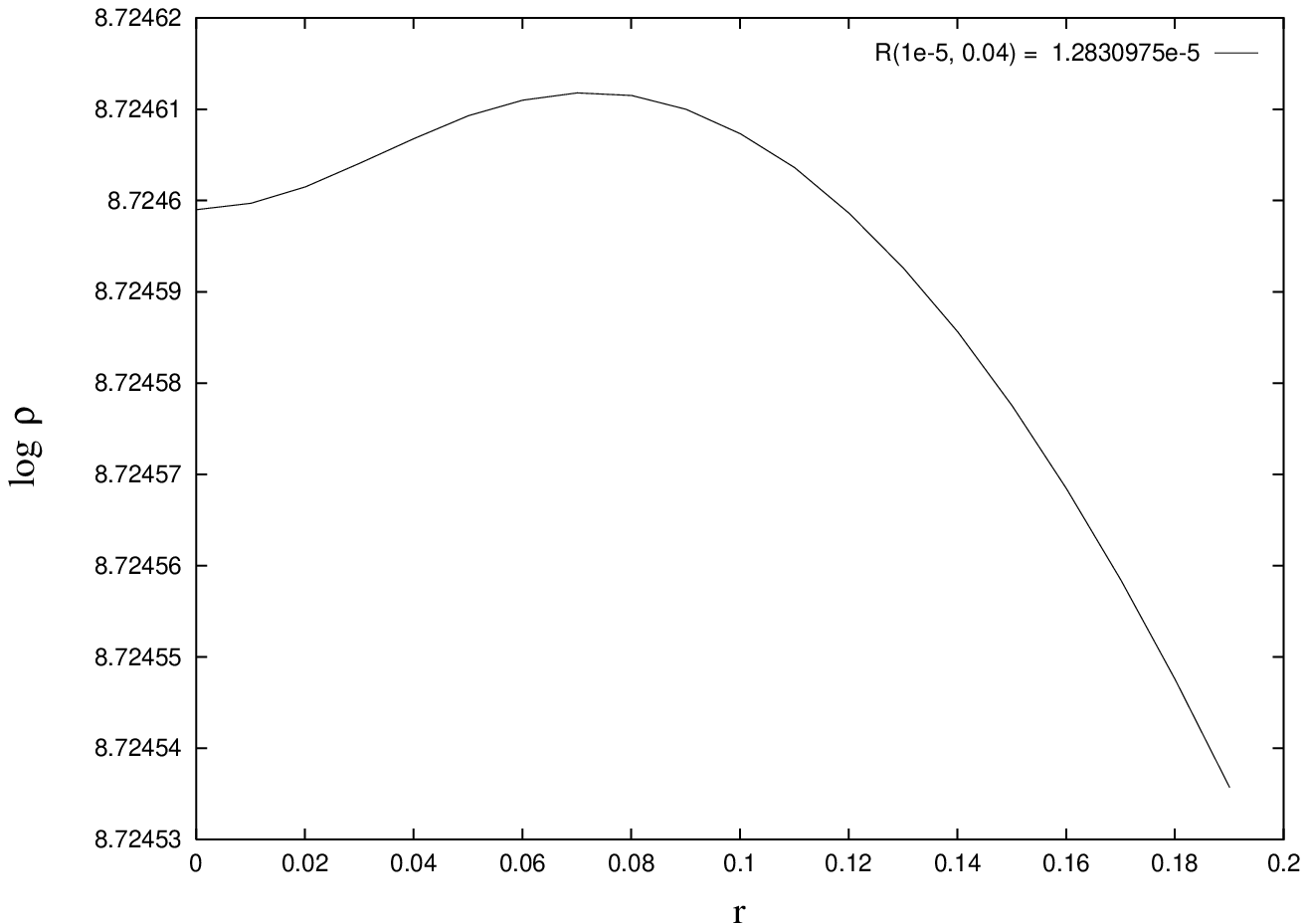}
 \caption{
 The density profile on a worldline at a later time.  This diagram and the
next one illustrate the change in concavity at the centre, which occurs
when the universe was $\approx 2 \times 10^4$ years old.  $R \approx
7.9$~kpc and $\rho_0 \approx 2.0 \times 10^{-18}$~g/cc.
 }
 \label{fi:dpi2}
 \end{figure}

 \begin{figure}[h]
 \epsffile[0 0 385 266]{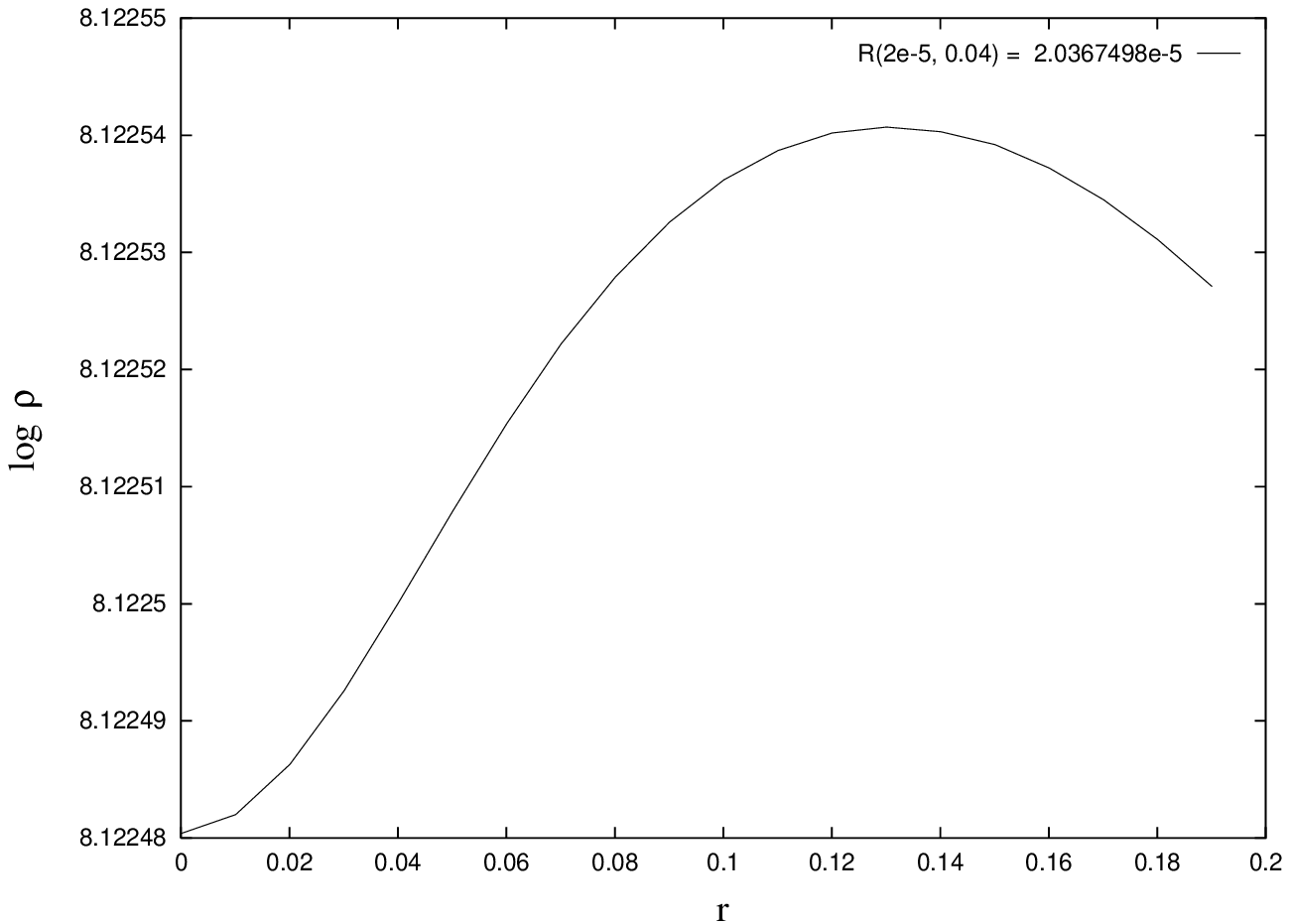}
 \caption{
 The density profile on a worldline at a still later time which, when
compared to the previous figure, illustrates the movement of the maximum
away from the centre.  This is as we expected and shows that a density
wave must exist near the origin if the profile inverts on the central
worldline.  $t \approx 4 \times 10^{4}$~yrs, $R \approx 12.5$~kpc and
$\rho_0 \approx 5 \times 10^{-19}$~g/cc.
 }
 \label{fi:dpi3}
 \end{figure}

 \begin{figure}[h]
 \epsffile[0 0 375 265]{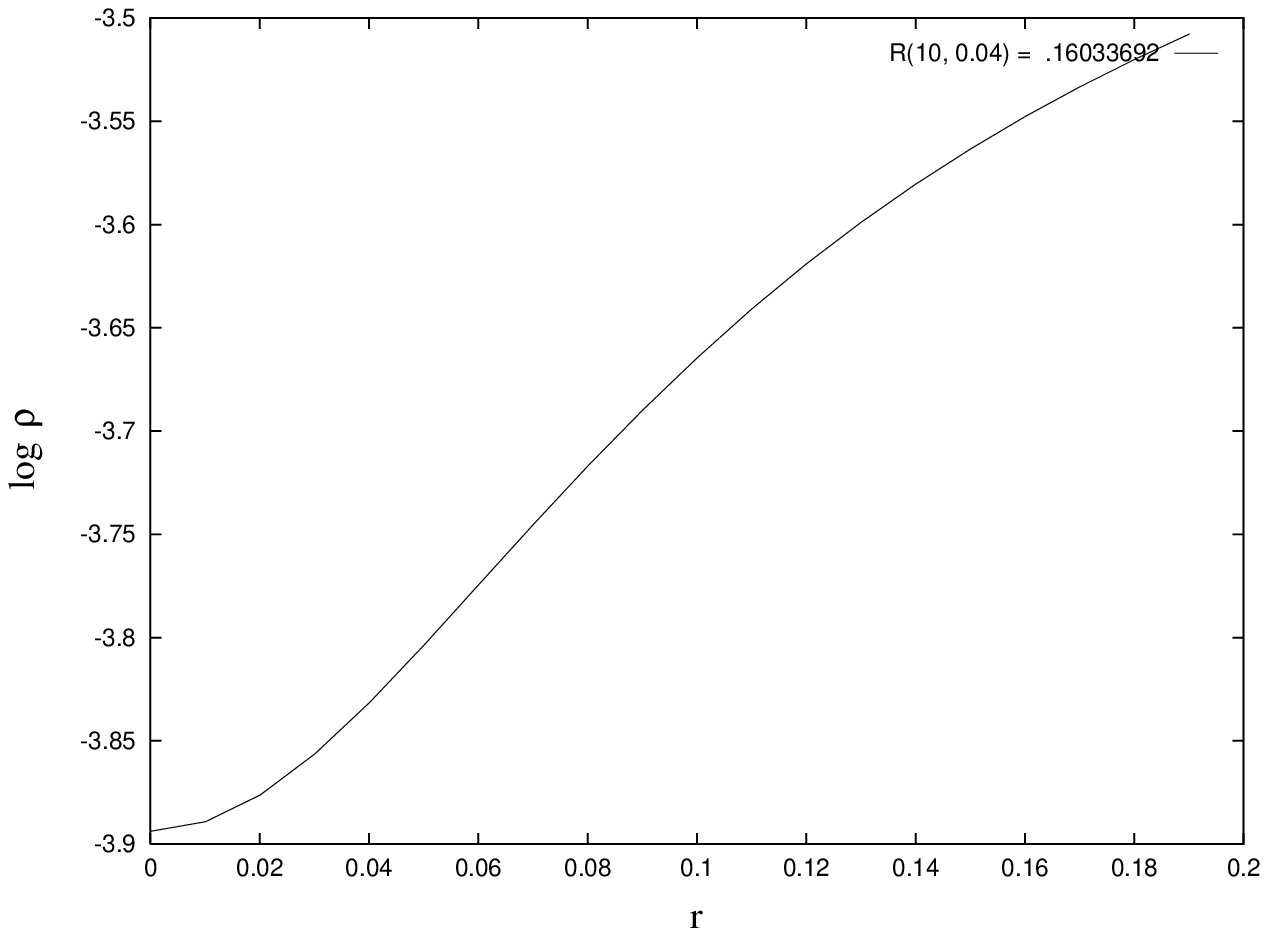}
 \caption{
 The density profile on the worldline today ($t \approx 20$~Gyrs).  This
corresponds to a void (albeit one with a rather elongated wall) with a
diameter of approximately $100$~Mpc.  $\rho_0 \approx 5 \times 10^{-
31}$~g/cc and the maximum density on this diagram is $\rho_m \approx 10^{-
3.5}$~cmu/clu$^3$ $\approx 1.2 \times 10^{-30}$~g/cc.
 }
 \label{fi:dpi4}
 \end{figure}

 \end{document}